\documentclass[aps,pre,superscriptaddress,floatfix,12pt]{revtex4-1}
\usepackage{amsmath,amsfonts,amssymb}
\usepackage{graphicx}
\usepackage{algorithm}
\usepackage{algpseudocode}
\usepackage{epstopdf}
\usepackage{float,placeins}
\usepackage[hyperfigures,breaklinks,colorlinks,citecolor=blue,linkcolor=black]{hyperref}
\usepackage[caption=false]{subfig}
\usepackage{mathbbol}

\def\s{\sigma}
\def\ul{\underline}
\def\p{\partial}
\def\partition{\mathcal{Z}}

\begin{document}
\title{The Cavity Master Equation: average and fixed point of the ferromagnetic model in random graphs}

\author{E. Dom\'{\i}nguez}
\affiliation{Group of Complex Systems and Statistical Physics. Department of Theoretical Physics, Physics Faculty, University of Havana, Cuba}
\email{eduardo@fisica.uh.cu}
\author{D. Machado}
\affiliation{Group of Complex Systems and Statistical Physics. Department of Theoretical Physics, Physics Faculty, University of Havana, Cuba}
\email{dmachado@fisica.uh.cu}
\author{R. Mulet}
\affiliation{Group of Complex Systems and Statistical Physics. Department of Theoretical Physics, Physics Faculty, University of Havana, Cuba}
\email{mulet@fisica.uh.cu}

\date{\today}

\begin{abstract}
\noindent
The Cavity Master Equation (CME) is a closure scheme to the usual Master Equation representing the dynamics of discrete variables in continuous time. In this work we explore the CME for a ferromagnetic model in a random graph. We first derive and average equation of the CME that describes the dynamics of mean magnetization of the system. We show that the numerical results compare remarkably well with the Monte Carlo simulations. Then, we show that the stationary state of the CME is well described by BP-like equations (independently of the dynamic rules that let the system towards the stationary state). These equations may be rewritten exaclty as the fixed point solutions of the Cavity Equation if one also assumes that the stationary state is well described by a Boltzmann distribution.

 
\end{abstract}
\maketitle

\section{Introduction}


The comprehension of non-equilibrium phenomena in complex systems is fundamentally more difficult than the understanding of their equilibrium properties for at least two reasons. First, because approximations which work for short time scales are not necessarily valid at long time scales and vice versa, hindering the possibility to have one solution able to describe the complete dynamical process. Second, because the dynamical rules that define the processes are  fundamental to characterize the evolution of the macroscopic quantities of the model \cite{gardiner1985handbook,van1992stochastic} and this opens a vast range of different dynamics for a model that are absent when we look only to its equilibrium properties. This is particularly unfortunate because many physical systems and all the living ones operate out-of-equilibrium.

A general approach to gain proper insight in these systems is to develop simple models that, although composed of many interacting particles, can be treated analytically and/or computationally on reasonable time scales. A proper classification of those models and the dynamical rules in terms of very general properties permitted the development of advanced techniques applicable to many problems within the same class.

A first classification of non-equilibrium processes can be made looking to the cardinality of the variables involved. They can be continuous or discrete. The dynamical modeling in the former case is usually done by using a Langevin-like equation \cite{van1992stochastic} where the time is also a continuous variable. If variables are discrete one describes the stochasticity of the dynamics by writing equations for the probability of the spin state. In this case, one chooses between two possible ways of describing the dynamics: either time evolves in discrete steps, or continuously.

In the case of a continuous-time a proper dynamic description of the spin state configuration, follows a Master Equation (ME) for the probability density of the states of $N$-spin interacting variables \cite{Glauber63, coolen2005theory}. Unfortunatelly the full solution of the master equation in the general case is a cumbersome task and exact solutions have been limited to simple models \cite{Glauber63,mayer2004general}. Alternatively the Dynamical Replica Analysis for fully connected \cite{laughton1996order} and diluted graphs \cite{Hatchett05,mozeika2008dynamical,barthel2015matrix} permits the derivation of  dynamical equations for the probability of some macroscopic observables. This approach obviously reduces the dimensionality of the problem, making it easir to treat, but looses detailed information about the microscopic state of the system. 

A more recent approach named the Cavity Master Equation (CME), provides a new method to close the Master Equation representing the continuous time dynamics of discrete interacting variables. The method makes use of the theory of Random Point Processes and has been succesfully applied to describe the dynamics of several models in graphs with finite connectivity, the Ising ferromagnet, the Random Field Ising model, and the Viana-Bray spin-glass model \cite{CME-PRE}, the ferromagnetic $p$-spin under Glauber dynamics \cite{CME-PSPIN} and more recently also the dynamics of a Focused Search algorithm to solve the 3-SAT problem in a random graph\cite{CME-PRL}.

In this work we continue to explore the CME. We start by introducing it once again, hopefully making easier for the reader to understand how it is different from previous approaches, and pointing its current limitations. We then focus our attention into Glauber dynamics of the ferromagnetic Ising model in a diluted graph with random connectivity, which is very well described by the CME \cite{CME-PRE}. In this context we derive an average case version of the CME, which has been absent in the literature, and we show that it provides a very good description of the average magnetization's dynamics. This equation is complementary to the more exact and cumbersome local description provided by the original CME. Finally we connect the long time dynamics of the CME with the celebrated equilibrium cavity method and Belief Propagation (BP) equations, extensively used to study the equilibrium properties of disordered systems in random graphs\cite{yedidia2005constructing,yedidia2001idiosyncratic,riccitersenghi2012,mezard2009information}. We hope that this connection could provide new insights and future developments in the comprehension of the CME, in particular in providing new ways to unveil the dynamics of disordered systems in the glassy phase\cite{bouchaud1998out}.



\section{The Model Dynamics}\label{sec:model}

Consider a system of $N$ interacting discrete variables $\ul\sigma = \{\sigma_1,\dots, \sigma_N\}$, with $\s_i = \pm 1$, with transition rate $r_i(\ul\sigma)$. The Master Equation (ME) describing the evolution of the probability of the system to be in state $\underline{\sigma}(t)$ at time $t$ is \cite{Glauber63, coolen2005theory}
\begin{equation}
\frac{d P (\ul\sigma) }{dt} = - \sum_{i=1}^N \Big[
r_i(\ul\sigma) P (\ul\sigma) - r_i(F_i(\ul\sigma)) P (F_i(\ul\sigma) ) \Big]\, ,
\label{eq:originalME}
\end{equation}
where we omitted the time dependence in $P(\ul\sigma,t)$ to shorten notation and $F_i$ represents the inversion operator on spin $i$, \emph{i.e.} $F_i (\ul\sigma) = \{\sigma_1, \dots, \sigma_{i-1}, -\sigma_{i},\sigma_{i+1}, \dots, \sigma_N\}$.

Although (\ref{eq:originalME}) is a simple equation to state formally, in practice it implies the daunting task of tracking the evolution in time of $2^N$ discrete states.  However, if $r_i(\ul\sigma)$ depends only on the configuration of spin $i$ and some neighbours $\partial i$, the master equation can be reduced to a local form. The evolution in time of the probability of the spin configuration $\sigma_i$ is then obtained by tracing (\ref{eq:originalME}) over all the spin states except $\sigma_i$. This Local Master Equation (LME) reads
\begin{equation}
\frac{d P (\sigma_i) }{dt} = - \sum_{\sigma_{\p i}}  \Big[
r_i(\sigma_i, \sigma_{\p i}) P (\sigma_i,\sigma_{\p i} ) -  r_i(-\sigma_i, \sigma_{\p i}) P (-\sigma_i,\sigma_{\p i} ) \Big]
\label{eq:localME}
\end{equation}
\noindent where $\sigma_{\p i}$ represents the configuration of all the spins in the neighbourhood of $i$. 
 
Contrary to \eqref{eq:originalME}, equation \eqref{eq:localME} is not closed. On the left hand side we have the probability $P(\sigma_i)$ that spin $i$ is in a particular state while on the right hand side,  $P(\sigma_i,\sigma_{\p i} )$ is the probability of a certain configuration for spin $i$ and its neighbours. To describe the evolution of the single site probability \eqref{eq:localME} in time, we then have to search for a closure of this equation. The simplest clousure scheme $P (\sigma_i,\sigma_{\p i} ) = \prod_{j \in {i,\p i}} P_j(\sigma_j)$ leads to the mean-field approximation, that although simplifies considerably the task is clearly wrong for diluted graphs. 

On the other hand, one may assume that 
\begin{equation}
P(\sigma_i,\sigma_{\p i} ) = \prod_{k \in \p i } P(\sigma_k | \sigma_i) P(\sigma_i)
\label{eq:factorP}
\end{equation}
\noindent which has the desirable property of being exact at equilibrium for trees and random graphs where loops are large compared to the system size.  Assuming a tree-like topology and the factorization in \eqref{eq:factorP}, the master equation (\ref{eq:localME}) can then be written as:
\begin{equation}
\frac{d P (\sigma_i) }{dt} = - \sum_{\sigma_{\p i}} \Big[
r_i(\sigma_i, \sigma_{\p i}) \big[ \prod_{k \in \p i }
P(\sigma_k| \sigma_i) \big] P(\sigma_i)
-  r_i(-\sigma_i, \sigma_{\p i}) 
\big[ \prod_{k \in \p i }
P(\sigma_k| -\sigma_i) \big]P(-\sigma_i) \Big]
\label{eq:localMEfact}
\end{equation}
The above equation is also not closed, as we do not know how $P(\sigma_k| \sigma_i)$ changes with time. The Cavity Master Equation is an equation for a cavity probability  $p(\s_i |\s_k)$ that approximates $P(\s_i|\s_k)$ and under certain conditions is equal to it. CME reads:

\begin{align}
 \nonumber
 \frac{d p(\s_i|\s_j)}{d t} = -  \sum_{\sigma_{\partial i\setminus j}} \Bigg[& r_{i}[\sigma_i,\sigma_{\partial i}]  \Big[\prod_{k\in\partial i \setminus j } p(\sigma_k|\s_i)\Big] p(\s_i|\s_j) \\
 &-  r_{i}[-\sigma_i,\sigma_{\partial i}] \Big[\prod_{k\in\partial i \setminus j } p(\sigma_k|-\s_i)\Big] p(-\s_i|\s_j) \Bigg]
\label{eq:CME}
\end{align}

This equation was derived in detail \cite{CME-PRE} starting from the  Random Point Process formalism where a specific spin history or trajectory $X$ is parametrized by the number of spin flips\cite{van1992stochastic,daley2007introduction,srinivasan1969stochastic}, the time in which they occur and the initial state of the system. It constitutes the starting point for the rest of this work.%

\section{The Average Case Cavity Master Equation}
In this section we will derive and numerically test an average case equation corresponding to CME in Erdos-Renyi graphs. The formalism developed with this purpose will be shown in (\ref{sec:ACME_form}), and a comparison with regular CME will be displayed in (\ref{sec:num_results}), alongside an explanation of what the average behavior is in Erdos-Renyi graphs.

\subsection{Formalism} \label{sec:ACME_form}

\begin{figure}[htb]
\centering
\includegraphics[keepaspectratio=true,width=0.4\textwidth]{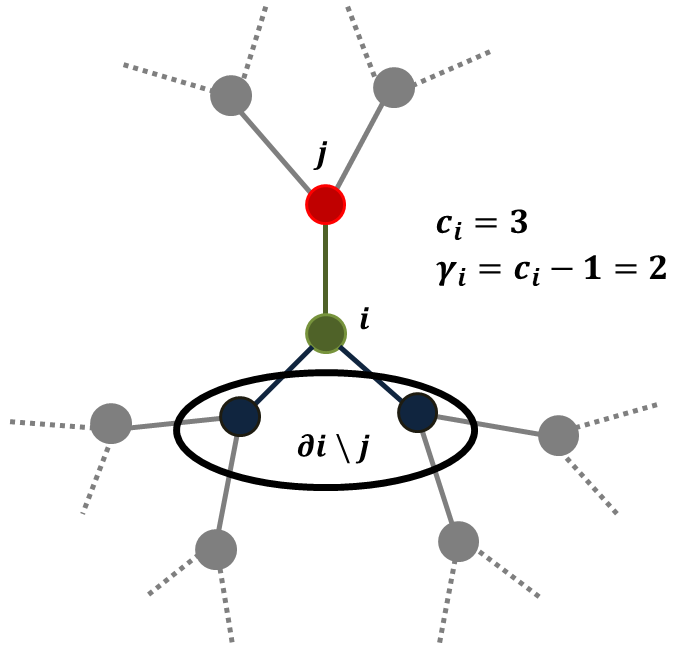}
\caption{Illustration of the local structure that determines the shapes of equations (\ref{eq:localMEfact}) and (\ref{eq:CME}). In this case node $i$ has connectivity $c_{i}=3$ and reduced connectivity $\gamma_{i} = 2$}
\label{fig:Illustration_ME_CME}
\end{figure}

When integrating CME (\ref{eq:CME}) and the Master Equation (\ref{eq:localMEfact}) on a given graph, we have a set of differential equations, one for each cavity probability of the form $p(\sigma_i \mid \sigma_j)$ nd one for each $P(\sigma_i)$. The number of terms in the sum that appears in CME (see (\ref{eq:CME})) is determined by the reduced connectivity $\gamma_i$ of node $i$ (an illustration is provided in figure (\ref{fig:Illustration_ME_CME})). Inside the sum, we have other cavity probabilities $p(\sigma_k \mid \sigma_i)$, which again follow similar equations whose shape depends on reduced connectivity $\gamma_k$. Again, in the case of Master Equation (\ref{eq:localMEfact}), the number of terms in the sum is determined by the connectivity $c_i$. In practice, the solution of equations (\ref{eq:CME}) becomes the proxy to solve the local master equation (\ref{eq:localMEfact}).


To find the equations that represent the average of (\ref{eq:localMEfact}) and (\ref{eq:CME}) over the full set of Erdos-Renyi graphs, we look first into the equations for a specific random graph, an instance of the ensemble. We can then try to approximately parametrize each equation by using the set $\lbrace \gamma_i,  \gamma_{k \in \partial i \setminus j} \rbrace$. We say that an instance of equation (\ref{eq:CME}) is determined by the set of reduced connectivities $\lbrace \gamma_i,  \gamma_{k \in \partial i \setminus j} \rbrace$ and the values of $\sigma_i$ and $\sigma_j$. Analogously, an instance of (\ref{eq:localMEfact}) will be determined by he set $\lbrace c_i,  \gamma_{k \in \partial i} \rbrace$, and by $\sigma_i$.

In an Erdos-Renyi's graph the probability of having a node with connectivity $c_i$ is given by $Q(c_i) = \frac{e^{- \langle c \rangle} \langle c \rangle^{c_i}}{c_i !}$, where $\langle c \rangle$ is the average connectivity. Then we get for the probability $q(\gamma_i)$ of having  node with reduced connectivity $\gamma_i$:

\begin{eqnarray}
&&q(\gamma_i) = \frac{(\gamma_i + 1) Q (c_i = \gamma_i + 1)}{\sum_{l=0}^{\infty} l \, Q(c_i = l)} = \frac{\gamma_i + 1}{\langle c \rangle} \, \frac{e^{- \langle c \rangle} \langle c \rangle^{\gamma_i + 1}}{(\gamma_i + 1) !} \nonumber \\
&&q(\gamma_i) = \frac{e^{- \langle c \rangle} \langle c \rangle^{\gamma_i}}{\gamma_i !} = Q(\gamma_i)
 \label{eqn:biased_connect}
\end{eqnarray}

This means that $\langle \gamma \rangle = \langle c \rangle \equiv \lambda$. The probability, for a given $\sigma_i$, of finding some instance of (\ref{eq:localMEfact}) is given by $Q(c_i) \, q(\gamma_{\partial i}) \equiv \prod_{k \in \partial i} q(\gamma_k) Q(c_i)$. Meanwhile, for some $(\sigma_i, \sigma_j)$, an instance of (\ref{eq:CME}) will appear with probability $q(\gamma_j, \gamma_{\partial j \setminus i}) \equiv \prod_{k \in \partial j \setminus i} q(\gamma_k) q(\gamma_j)$.

Now we will average equations (\ref{eq:localMEfact}) and (\ref{eq:CME}) using these probabilities. This procedure can be tested through a numerical scheme that samples different instances of the equations using $q(\gamma)$ and $Q(c)$, and the results justify continuing with our analytical procedure (see Appendix A). Taking, as it was said before, Glauber rules for the dynamics \cite{Glauber63}, multiplying by $\prod_{k \in \partial i} q(\gamma_k) Q(c_i)$ and summing over all $\lbrace c_i,  \gamma_{k \in \partial i} \rbrace$, we get a transformed Master Equation:

\begin{eqnarray}
\nonumber
 \lefteqn{\dot{P}_{\lambda}(\sigma_{i})=- \dfrac{\alpha}{2} \lbrace P_{\lambda}(\sigma_{i}) -  P_{\lambda}(-\sigma_{i}) \rbrace + \dfrac{\alpha}{2} \sigma_i \sum_{c_i=0}^{\infty} Q(c_i) \sum_{\sigma_{\partial i}} \tanh(\beta J \sum_{k \in \partial i} \sigma_k) \times }\\ 
 && \times \left[ \displaystyle \prod_{k \in \partial i} \left( \sum_{\gamma_pdfk=0}^{\infty} q(\gamma_k) p_{\gamma_k}( \sigma_{k} \mid \sigma_{i}) \right) \; P_{c_i}(\sigma_{i}) + \displaystyle \prod_{k \in \partial i} \left( \sum_{\gamma_k=0}^{\infty}  q(\gamma_k) p_{\gamma_k}( \sigma_{k}  \mid -\sigma_{i}) \right) \; P_{c_i}(-\sigma_{i})  \right] 
\label{eq:first_sum_ME_av_graph_1}
\end{eqnarray} 
where we defined $\sum_{\gamma_k=0}^{\infty} q(\gamma_k) p_{\gamma_k}( \sigma_{k} \mid \sigma_{i}) \equiv p_{\lambda}( \sigma_{k} \mid \sigma_{i})$ and $P_{\lambda} (\sigma') = \sum_c Q(c) P_c(\sigma')$ as average probabilities. Using that definition in the parenthesis of second line of (\ref{eq:first_sum_ME_av_graph_1}) we get:

\begin{eqnarray}
 \nonumber
 \lefteqn{\dot{P}_{\lambda}(\sigma_{i})=- \dfrac{\alpha}{2} \lbrace P_{\lambda}(\sigma_{i}) -  P_{\lambda}(-\sigma_{i}) \rbrace + \dfrac{\alpha}{2} \sigma_i \sum_{c_i=0}^{\infty} Q(c_i) \sum_{\sigma_{\partial i}} \tanh(\beta J \sum_{k \in \partial i} \sigma_k) \times }\\ 
 && \times \left[ \displaystyle \prod_{k \in \partial i} p_{\lambda}( \sigma_{k} \mid \sigma_{i}) \; P_{c_i}(\sigma_{i}) + \displaystyle \prod_{k \in \partial i} p_{\lambda}( \sigma_{k}  \mid -\sigma_{i})  \; P_{c_i}(-\sigma_{i})  \right] 
\label{eq:first_sum_ME_av_graph_2}
\end{eqnarray} 

Similarly, multiplying by $\prod_{k \in \partial i \setminus j} q(\gamma_k) q(\gamma_i)$ and summing over all $\lbrace \gamma_i,  \gamma_{k \in \partial i \setminus j} \rbrace$, we get a transformed Cavity Master Equation:

 \begin{eqnarray}
 \nonumber
 \lefteqn{\dot{p}_{\lambda}(\sigma_{i}| \sigma_j)=- \dfrac{\alpha}{2} \lbrace p_{\lambda}(\sigma_{i} \mid \sigma_{i}) -  p_{\lambda}(-\sigma_{i} \mid \sigma_{j}) \rbrace + \dfrac{\alpha}{2} \sigma_i \sum_{\gamma_j=0}^{\infty} q(\gamma_i) \sum_{\sigma_{\partial j \setminus i}} \tanh(\beta J \sum_{k \in \partial i} \sigma_k) \times }\\ 
 && \times \left[ \displaystyle \prod_{k \in \partial i \setminus j} p_{\lambda}( \sigma_{k} \mid \sigma_{i}) \; p_{\gamma_i}(\sigma_{i} \mid \sigma_{j}) + \displaystyle \prod_{k \in \partial i \setminus j} p_{\lambda}( \sigma_{k} \mid - \sigma_{i}) \; p_{\gamma_i}(-\sigma_{i} \mid \sigma_{j})  \right]
\label{eq:first_sum_CME_av_graph_2}
\end{eqnarray}

A numerical scheme can also be designed to test the validity of equations (\ref{eq:first_sum_ME_av_graph_2}) and (\ref{eq:first_sum_CME_av_graph_2}) (see Appendix A).

Now we must work on the infinite sums in the equations. Let's rewrite (\ref{eq:first_sum_ME_av_graph_2}) in therms of the average magnetization $m_{\lambda} = \sum_{\sigma'} \sigma' \, P_{\lambda}(\sigma')$. This is done by applying the $\sum_{\sigma_i} \sigma_i \langle \cdot \rangle$ operator to the equation, and we get:

\begin{eqnarray}
\dot{m}_{\lambda} =- \alpha m_{\lambda} + \alpha \sum_{\sigma'} \sum_{c=0}^{\infty} Q(c) \, P_{c}(\sigma') \sum_{\sigma_{\partial i}} \tanh(\beta J \sum_{k \in \partial i} \sigma_k) \prod_{k \in \partial i} p_{\lambda}(\sigma_k \mid \sigma')
 \label{eq:MEAv_RGER_m_eq_1}
\end{eqnarray}
where we have used that, after multiplying by $\sigma_i$ an summing over the same variable, the equation is not longer indexed in $\sigma_i$, and that the expression inside the sum in the second term of (\ref{eq:first_sum_ME_av_graph_2}) is even under $\sigma_i \rightarrow -\sigma_i$. We can then remove the factor $2$ from the denominator in the second term and shorten the equation by introducing a sum over a variable $\sigma'$ that takes the place of the former $\sigma_i$. We also re-denoted $c_i$ into $c$ for simplicity.

The infinite sum in (\ref{eq:MEAv_RGER_m_eq_1}) can be approximated using an analogy with \\ $\int_{m=-1}^{1} \tanh(m) P(m) \;dm \approx \tanh \left( \int_{m=-1}^{1} m P(m) \;dm \right)$ to get:

\begin{eqnarray}
\dot{m}_{\lambda} &=&- \alpha m_{\lambda} + \alpha \sum_{\sigma'} \sum_{c=0}^{\infty} Q(c) \, P_{c}(\sigma')  \tanh \left( \beta J \sum_{\sigma_{\partial i}} \sum_{k \in \partial i} \sigma_k \prod_{l \in \partial i} p_{\lambda}(\sigma_l \mid \sigma') \right) \nonumber \\
&=&- \alpha m_{\lambda} + \alpha \sum_{\sigma'} \sum_{c=0}^{\infty} Q(c) \, P_{c}(\sigma')  \tanh \left( \beta J \sum_{k \in \partial i} \left[ \prod_{l \in \partial i \setminus k} \sum_{\sigma_{l}} p_{\lambda}(\sigma_l \mid \sigma') \right] \left[ \sum_{\sigma_k}  p_{\lambda}(\sigma_k \mid \sigma')  \sigma_k \right]  \right) \nonumber \\
&=&- \alpha m_{\lambda} + \alpha \sum_{\sigma'} \sum_{c=0}^{\infty} Q(c) \, P_{c}(\sigma')  \tanh \left( \beta J \sum_{k \in \partial i} \hat{m}_{\lambda}(\sigma')  \right) \nonumber \\
&=&- \alpha m_{\lambda} + \alpha \sum_{\sigma'} \sum_{c=0}^{\infty} Q(c) \, P_{c}(\sigma')  \tanh \left( \beta J \, c \, \hat{m}_{\lambda}(\sigma')  \right)
 \label{eq:MEAv_RGER_m_eq_tanh_outside_1}
\end{eqnarray}

From second line to third line of (\ref{eq:MEAv_RGER_m_eq_tanh_outside_1}) we have used the normalization  $\sum_{\sigma_{l}} p_{\lambda}(\sigma_l \mid \sigma') = 1$. We also have defined the cavity magnetization: \\

\begin{equation}
\hat{m}_{\lambda}(\sigma') \equiv \sum_{\sigma_k}  p_{\lambda}(\sigma_k \mid \sigma')  \sigma_k
\label{eq:cav_mag}
\end{equation}

We can similarly write an equation equivalent to (\ref{eq:MEAv_RGER_m_eq_tanh_outside_1}) but for $\hat{m}_{\lambda}(\sigma)$, which reads:

\begin{equation}
\frac{d \hat{m}_{\lambda}(\sigma)}{dt}=- \alpha \hat{m}_{\lambda}(\sigma) + \alpha \sum_{\sigma'} \sum_{\gamma=0}^{\infty} q(\gamma) \, p_{\gamma}(\sigma' \mid \sigma) \tanh \left[\beta J \gamma \, \hat{m}_{\lambda}(\sigma')  + \beta  J \sigma \right]
 \label{eq:CMEAv_RGER_m_eq_tanh_outside_1}
\end{equation}

A final approximation in equations (\ref{eq:MEAv_RGER_m_eq_tanh_outside_1}) and (\ref{eq:CMEAv_RGER_m_eq_tanh_outside_1}) gives a closure and allows to express magnetization's dynamics in terms of differential equations for $m_{\lambda}$ and $\hat{m}_{\lambda}(\sigma)$ alone. Let's write it for equation (\ref{eq:MEAv_RGER_m_eq_tanh_outside_1}):

\begin{eqnarray}
\dot{m}_{\lambda} &=&- \alpha m_{\lambda} + \alpha \sum_{\sigma'} \sum_{c=0}^{\infty} Q(c) \, P_{c}(\sigma')  \tanh \left( \beta J \, c \, \hat{m}_{\lambda}(\sigma')  \right) \nonumber \\
&=& - \alpha m_{\lambda} + \alpha \sum_{\sigma'} \left( \sum_{c=0}^{\infty} Q(c) \, P_{c}(\sigma') \right) \left( \sum_{c=0}^{\infty} Q(c)  \tanh \left[ \beta J \, c \, \hat{m}_{\lambda}(\sigma')  \right] \right) \nonumber \\
&=& - \alpha m_{\lambda} + \alpha \sum_{\sigma'} P_{\lambda}(\sigma') \sum_{c=0}^{\infty} Q(c)  \tanh \left[ \beta J \, c \, \hat{m}_{\lambda}(\sigma')  \right] \nonumber \\
&=& - \alpha m_{\lambda} + \alpha \sum_{\sigma'} \frac {1 + \sigma' m_{\lambda}}{2} \sum_{c=0}^{\infty} Q(c)  \tanh \left[ \beta J \, c \, \hat{m}_{\lambda}(\sigma')  \right]
 \label{eq:MEAv_RGER_trick_derivative_last}
\end{eqnarray}

In (\ref{eq:MEAv_RGER_trick_derivative_last}) we have independently computed the averages of $P_c(\sigma')$ and $\tanh \left( \beta J \, c \, \hat{m}_{\lambda}(\sigma')  \right)$, $i.e$, we have assumed that the average of the product that appears in the first line is well approximated by the product of the averages in the second line. We also used that $P_{\lambda}(\sigma') = (1 + \sigma' m_{\lambda}) / 2$.

The equivalent equation for $\hat{m}_{\lambda}(\sigma)$ is:

\begin{equation}
\frac{d \hat{m}_{\lambda}(\sigma)}{dt} = - \alpha \hat{m}_{\lambda}(\sigma) + \alpha \sum_{\sigma'} \frac {1 + \sigma' \hat{m}_{\lambda}(\sigma)}{2} \sum_{\gamma=0}^{\infty} q(\gamma)  \tanh \left[\beta J \gamma \, \hat{m}_{\lambda}(\sigma')  + \beta  J \sigma \right]
 \label{eq:CMEAv_RGER_trick_derivative_last}
\end{equation}

Equations (\ref{eq:MEAv_RGER_trick_derivative_last}) and (\ref{eq:CMEAv_RGER_trick_derivative_last}) can be numerically integrated. In the following subsection we will compare the results with the ones of equations (\ref{eq:localMEfact}) and (\ref{eq:CME}).

\subsection{Numerical results} \label{sec:num_results}

Let's start by showing what an average over Erdos-Renyi graphs means for equations (\ref{eq:localMEfact}) and (\ref{eq:CME}). We will study the dynamics of the Ising ferromagnet, whose well-known Hamiltonian is:

\begin{equation}
H = -J \sum_{\langle i j \rangle} \sigma_i \sigma_j
 \label{eq:Ising_ferromagnet_Hamiltonian}
\end{equation}
where $J$ is a coupling constant that we will take equal to one, and $\langle i j \rangle$ are the indexes of all interacting pairs in the system.

\begin{figure}[htb]
\centering
\includegraphics[keepaspectratio=true,width=0.45\textwidth]{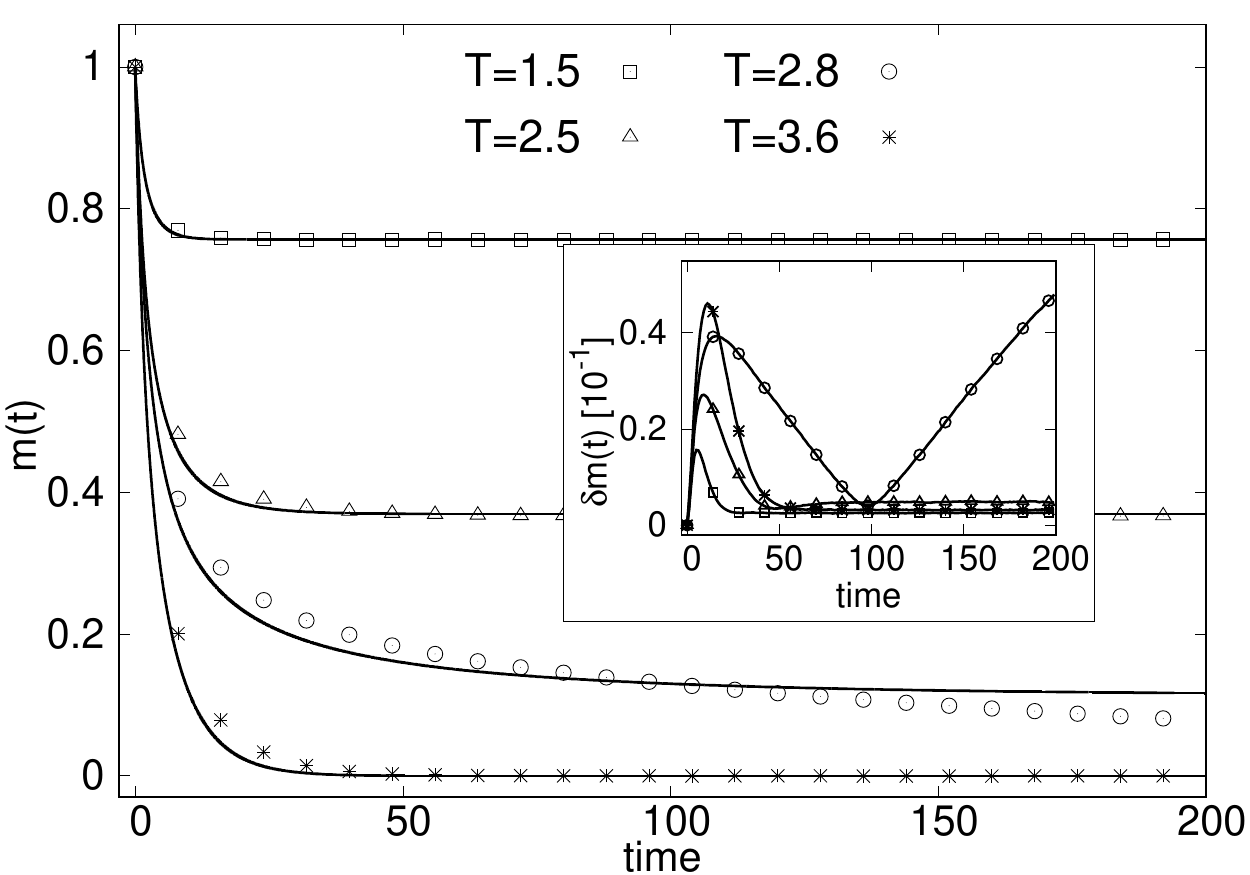}
\includegraphics[keepaspectratio=true,width=0.45\textwidth]{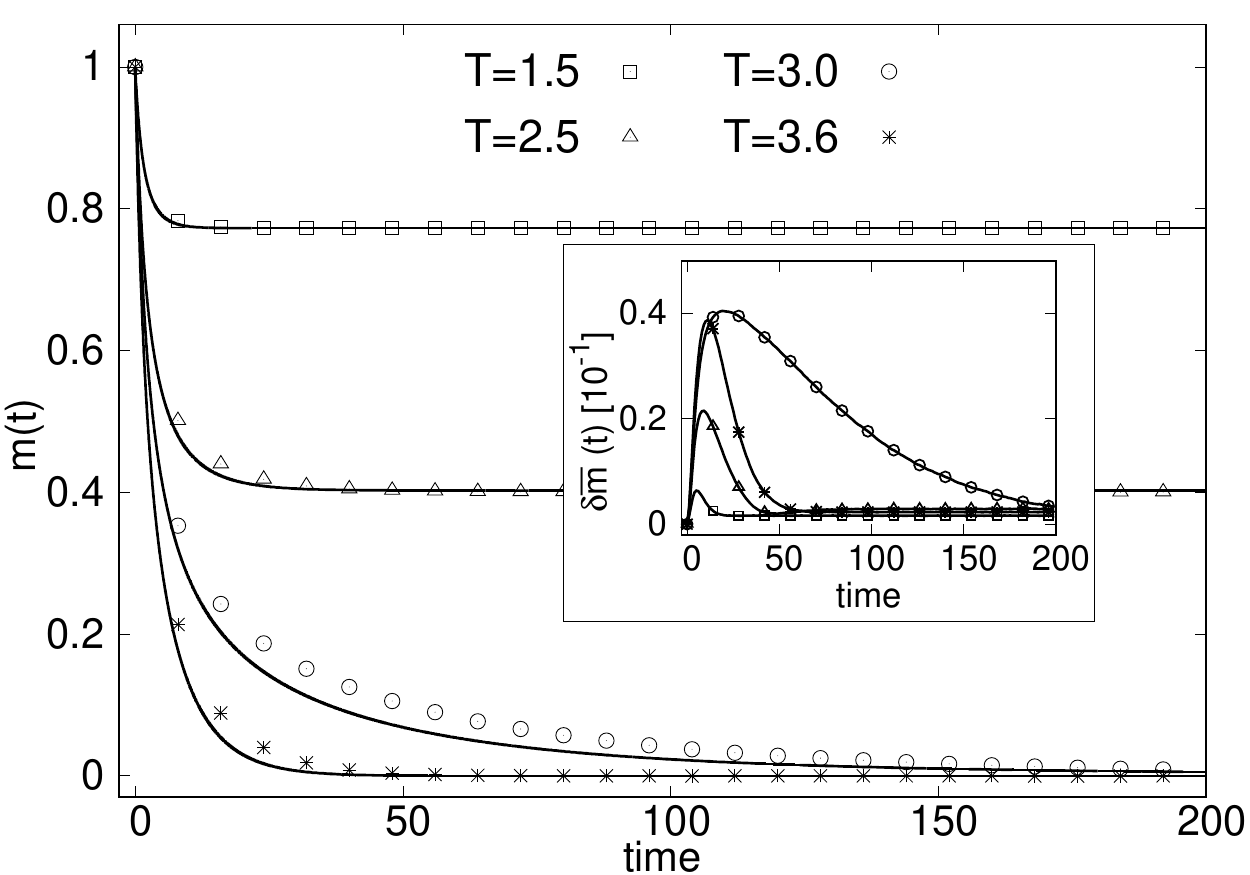}
\caption{Comparison between CME's (lines) and Monte Carlo's (points) results in\
 Erdos-Renyi graphs with mean connectivity $\langle c \rangle = 3$ ($T_c^{ER} \\
approx 2.89$). All calculations began with a fully magnetized system in contact\
 with a heath bath at a given temperature $T$. Main panels of both graphics sho\
w the time evolution of magnetization for several temperatures, and insets show\
 the correspondent local error $\delta m$ (see equation (\ref{eq:local_error}))\
 or $\delta \overline{m}(t)$ (see equation (\ref{eq:local_error_average})). Sys\
tem's size is in all cases $N=4000$. (left) Single instance. Each point represe\
nts the average over $n=100000$ Monte Carlo's histories, and each line correspo\
nd to a single integration of CME, for only one Erdos-Renyi graph. (right) Aver\
age over graphs. Each point is the result of two consecutive averages, the firs\
t one over $n=100000$ Monte Carlo's histories, and the second one over $s=350$ \
different graphs. CME's lines represent are also an average over $s=350$ graphs\
. Error bars are in all cases of the size of the points in the figure.}
\label{fig:CME_vs_MC_ER_graphs}
\end{figure}

With the model already defined, only the structure of the interactions, $i.e.$, the interactions graph, is necessary to obtain numerical results from (\ref{eq:localMEfact}) and (\ref{eq:CME}). Figure (\ref{fig:CME_vs_MC_ER_graphs}) shows a comparison between the integration of these equations and Monte Carlo simulations for Erdos-Renyi graphs. The main panel in (\ref{fig:CME_vs_MC_ER_graphs} (left)) shows the time evolution of system's magnetization for several temperatures, and the inset displays the corresponding behavior of the local error:

\begin{equation}
\delta m(t) = \sqrt{\frac{1}{N} \sum_{i} \left( m_{i}^{CME}(t) - m_{i}^{MC}(t) \right)^{2}}
 \label{eq:local_error}
\end{equation}

Our CME method accurately reproduces the output of Monte Carlo simulations. As can be seen in main panel of figure (\ref{fig:CME_vs_MC_ER_graphs} (left)), which was done for graphs with mean connectivity $\langle c = 3 \rangle$, the steady-state value of Monte Carlo's magnetization is well described by CME in all cases, except for $T=2.8$. The latter is close to model's critical temperature $T_c^{ER} = 1 / \text{arctanh}(1 / 3)$. Transient regime is also reproduced, and the similarity increases when we move away from critical behavior, both below and above $T_c^{ER}$.

Each Monte Carlo's point in the figure represents two consecutive averages. For a given graph, and always starting from a fully magnetized system in contact with a heat bath, we averaged several Monte Carlo's histories. We repeated that procedure for some set of Erdos-Renyi graphs with the same mean connectivity, and averaged the outputs. That last average is what is shown in figure (\ref{fig:CME_vs_MC_ER_graphs} (right)). On the other hand, CME itself is written in the language of probabilities, so the first average that we did with Monte Carlo is not needed. Nevertheless, we performed the second average over different graphs in the same way to get the results in figure (\ref{fig:CME_vs_MC_ER_graphs} (right)).

The local error shown in the inset of figure (\ref{fig:CME_vs_MC_ER_graphs} (left)) correspondingly haves a maximum at the transient, and then goes to a steady-state value for long times at temperatures which are not too close to $T_c^{ER}$. The maxima are higher near $T_c^{ER}$, and the steady-state value is close to zero. Near $T_c^{ER}$ (at $T=2.8$), error is large for short and long times. Problems related to criticality are a direct consequence of the approximations made in the derivation of the Cavity Master Equation \cite{CME-PRE}, and in the assumption given by equation (\ref{eq:factorP}).

In order to compare with the average case equations (\ref{eq:MEAv_RGER_trick_derivative_last}) and (\ref{eq:CMEAv_RGER_trick_derivative_last}) we need to study an average over several graphs of single instance results like the ones in (\ref{fig:CME_vs_MC_ER_graphs} (left)). In figure (\ref{fig:CME_vs_MC_ER_graphs} (right)) we averaged CME's and Monte Carlo's results over graphs with mean connectivity $\langle c \rangle = 3$. By doing so, we sample the mean behavior over the graphs distribution.

Let's clarify that local error shown in figure (\ref{fig:CME_vs_MC_ER_graphs} (right)) was actually computed with the average local magnetizations, it is not the average of local errors computed with local magnetizations as in (\ref{eq:local_error}). The formula would be now:

\begin{equation}
\delta \overline{m}(t) = \sqrt{\frac{1}{N} \sum_{i} \left( \langle m_{i}^{CME}(t) \rangle - \langle m_{i}^{MC}(t) \rangle \right)^{2}}
 \label{eq:local_error_average}
\end{equation}
where we have denoted the average over graphs as $\langle \cdot \rangle$. The reason why we choose to use (\ref{eq:local_error_average}) is that computing the error for a given graph and then averaging over graphs is more expensive from a computational point of view.

Due to fluctuations, computing with some confidence the average over different graphs near $T_c^{ER}$ is also computationally difficult. A large number of graphs is needed. For the purposes of this section and in view of the results illustrated in figure (\ref{fig:CME_vs_MC_ER_graphs} (left)), it is enough to show CME's average behavior for temperatures below and above $T_c^{ER}$. Steady-state and transient are well described, and the inset shows that local error has a maximum in transient, and then goes to a small value for long times. Maxima are higher when the temperature is close to $T_c^{ER}$.

The shape of equation (\ref{eq:local_error_average}) explains why steady-state error is significantly smaller in figure (\ref{fig:CME_vs_MC_ER_graphs} (right)) than in figure (\ref{fig:CME_vs_MC_ER_graphs} (left)). Although for some graphs CME's magnetization is smaller than Monte Carlo's one, for others this relation is inverted. In average, these two different behaviors cancel, the local magnetizations inside the sum in (\ref{eq:local_error_average}) are close to each other, and error is smaller. We should remark here that, in obtaining these results, a typical computation with Monte Carlo takes considerably more time than with the CME method.

\begin{figure}[htb]
\centering
\includegraphics[keepaspectratio=true,width=0.55\textwidth]{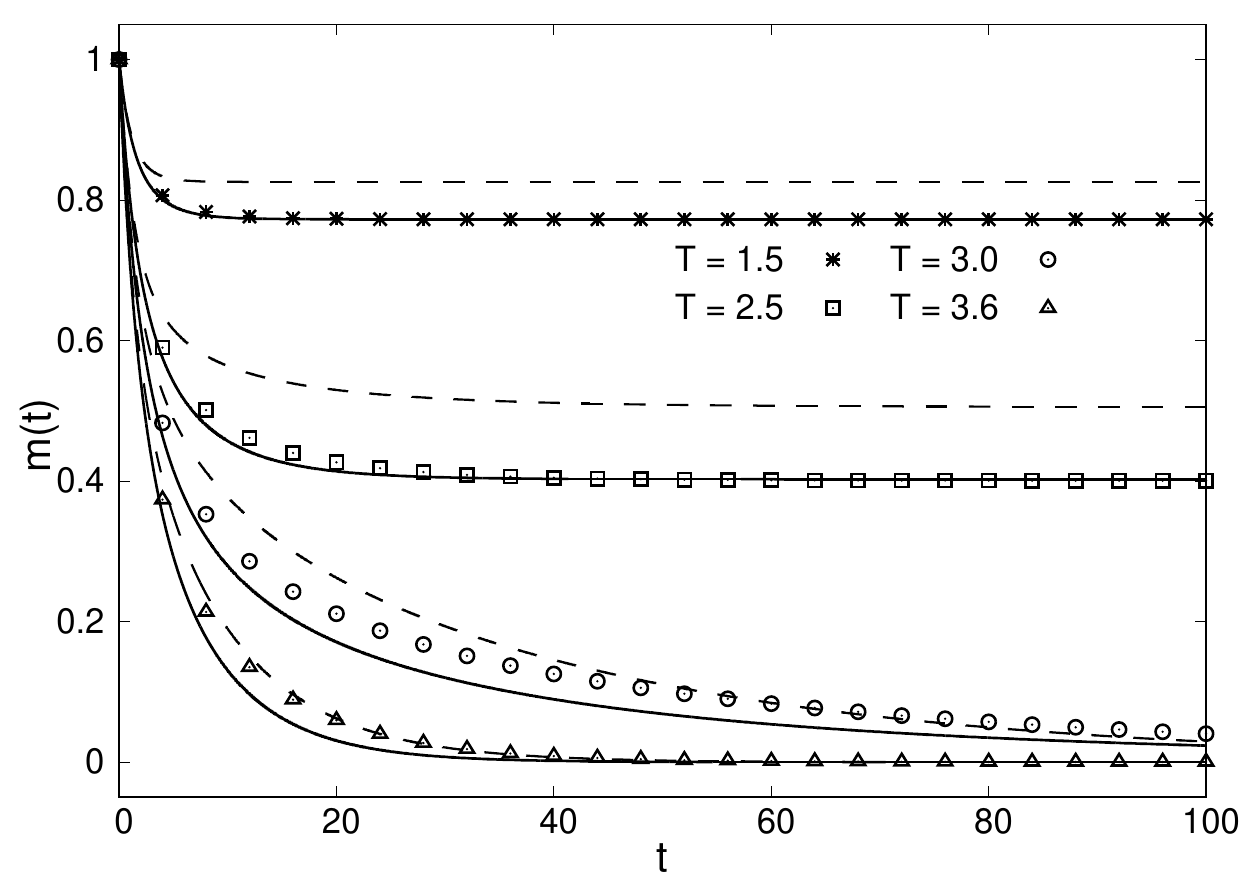}

\caption{Average CME's calculations (dashed lines) with initial magnetization $m_0=1.0$ in an Erdos-Renyi graph whose mean connectivity is $\langle c \rangle =3$ ($T_c \approx 2.88$). In both graphics the sums in (\ref{eq:MEAv_RGER_trick_derivative_last}) and (\ref{eq:CMEAv_RGER_trick_derivative_last}) were approximately computed using the first $50$ terms. All calculations began with a fully magnetized system in contact with a heath bath at a given temperature $T$. System's size is in all cases $N=4000$. Continuous curves are an average over different graphs of the integration of equations (\ref{eq:localMEfact}) and (\ref{eq:CME}). The number of graphs was $s=350$. Points are Monte Carlo's results. In each one $n=100000$ MC's histories were averaged. Each dot represents also the average over different graphs. Error bars are in all cases of the size of the points in the figure.}
\label{fig:average_CME_tanh_trick_2}
\end{figure}

Now, we can compare with the average case equations. We can cut the sums in (\ref{eq:MEAv_RGER_trick_derivative_last}) and (\ref{eq:CMEAv_RGER_trick_derivative_last}) at some value of $c$ and $\gamma$, respectively, and obtain numerical results. Figure (\ref{fig:average_CME_tanh_trick_2}) shows a comparison with regular CME and Monte Carlo simulations. Both CME and Monte Carlo results are an average over a set of Erdos-Renyi graphs with mean connectivity $\langle c \rangle =3$.

The Figure (\ref{fig:average_CME_tanh_trick_2}) clearly shows that the predictions of the average equations describes very well the high temperature phase, it understimates sligthly the dynamics at low temperatures. We think that the main responsabile of this underestimation is the approximation:  $\int_{m=-1}^{1} \tanh(m) P(m) \;dm \approx \tanh \left( \int_{m=-1}^{1} m P(m) \;dm \right)$. The errors are large only near  $T_c$, where this approximation is not valid, and in addition where finite size effects and long range correlation are known to be more important.

\section{Long time dynamics, Belief Propagation and Cavity Equations}

The Cavity Master Equation (\ref{eq:CME}) is the result of properly differentiating a set of dynamic cavity message passing equations \cite{CME-PRE, del2015dynamic} that have been derived recently. They are an extension to continuous time dynamics of the cavity method, which has become the method of choice to solve the statics of models on diluted graphs. It is not clear though how the stationary
state reached by CME is related to the equilibrium results obtained from the original cavity method or the equivalent Belief Propagation (BP) algorithm. In this section we prove that indeed this connection exists, that the equilibrium solution to the CME in a tree-like topology corresponds to the exact conditional probabilities determined by the cavity solution.

\subsection{The CME and BP-like equations in stationary states}

Let's start by recalling that in the CME we work with conditional distributions $p(\s_i|\s_j)$, whereas the cavity method deals with distributions $\mu_{i\rightarrow j}(\s_i)$ of a different nature. The CME probability gives the probability of state $\s_i$ knowing the state of its neighbor
$\s_j$. The cavity $\mu_{i\rightarrow j}(\s_i)$, on the other hand, represents the distribution
of $\s_i$ in an alternate construction where the interaction with spin $j$ had been suppressed.

The solution of the cavity update equations
\begin{equation}
\begin{array}{ccl}
\mu_{i\rightarrow j}(\sigma_i) &=& \dfrac{1}{\partition_{i\rightarrow j}} \sum_{\{\s_k\}\setminus \s_j} \exp[\beta \s_i(h_i + \sum_{k\in \partial i\setminus j} J_{ik}\s_k)]\prod_{k\in \p i\setminus j} \mu_{k\rightarrow i}(\s_k)
 \end{array}
 \label{eqn:exact_cavity}
\end{equation}
is, for a diluted network, asymptotically exact in the system size. For the sake of having simpler expressions we define $m_{k\rightarrow i}(J_{ik},\s_i) = \sum_{\s_k}\exp [\beta J_{ik}\s_i \s_k]\mu_{k\rightarrow i}(\s_k)$, which will be quickly identified by the reader familiar with BP as the pair to site messages. The exact cavity (BP) recursion reads then \cite{jordan1999introduction}:
\begin{equation}
 \mu_{i\rightarrow j}(\sigma_i) = \dfrac{1}{\partition_{i\rightarrow j}} \exp[\beta h_{i}\s_i]\prod_{k\in \p i\setminus j} m_{k\rightarrow i}(J_{ik},\s_i)
 \label{eqn:exact_cavity_2}
\end{equation}

On the other hand, the stationarity condition for the CME $\frac{dp(\s_i|\s_j)}{dt} = 0$ produces a set of
coupled equations for the conditional probabilities very similar in structure to \eqref{eqn:exact_cavity}. Stationarity implies balancing transitions in and out
each state:
\begin{equation}
\sum_{\sigma_{\partial i\setminus j}} r_{i}(\sigma_i,\sigma_{\partial i}) \, p(\s_i|\s_j) \prod_{k\in\partial i \setminus j } p(\sigma_k|\s_i) = \sum_{\sigma_{\partial i\setminus j}} r_{i}(-\sigma_i,\sigma_{\partial i}) \, p(-\s_i|\s_j) \prod_{k\in\partial i \setminus j } p(\sigma_k|-\s_i) 
 \label{eq:stationarity_CME_full_main_text}
\end{equation}
By using that $p(-\sigma_i \mid \sigma_{j}) = 1 - p(\sigma_i \mid \sigma_{j})$, we get:
\begin{eqnarray}
p(\sigma_i \mid \sigma_j) &=&  \frac{\sum_{\sigma_{\partial i\setminus j}} r_{i}(-\sigma_i,\sigma_{\partial i}) \, \prod_{k\in\partial i \setminus j } p(\sigma_k|-\s_i)}{\sum_{\sigma_i} \sum_{\sigma_{\partial i\setminus j}} r_{i}(\sigma_i,\sigma_{\partial i}) \, \prod_{k\in\partial i \setminus j } p(\sigma_k|\s_i)}
 \label{eq:update_CME_numeric_equality}
\end{eqnarray}

The term in the denominator of (\ref{eq:update_CME_numeric_equality}) is the same for $p(\sigma_i \mid \sigma_j)$ and $p(-\sigma_i \mid \sigma_j)$, and we can identify it as a normalization factor. That allows us to write:

\begin{eqnarray}
p(\sigma_i \mid \sigma_j) &\propto&  \sum_{\sigma_{\partial i\setminus j}} r_{i}(-\sigma_i,\sigma_{\partial i}) \, \prod_{k\in\partial i \setminus j } p(\sigma_k|-\s_i) \nonumber \\
&\propto& \sum_{\sigma_{\partial i\setminus j}} \frac{\exp \left[ -\beta \, \sigma_i \left( \sum_{k \in \partial i} J_{ki} \sigma_k + h_i \right) \right]}{ \cosh \left[ \beta \left( \sum_{k \in \partial i} J_{ki} \sigma_k + h_i \right) \right]} \, \prod_{k\in\partial i \setminus j } p(\sigma_k|-\s_i)
 \label{eq:update_CME_numeric}
\end{eqnarray}

Indeed, rules (\ref{eqn:exact_cavity}) and (\ref{eq:update_CME_numeric}) have a similar shape. We can also derive the following equation for $P(\sigma_i)$ starting from the stationarity condition for equation (\ref{eq:localMEfact}):

\begin{eqnarray}
P(\sigma_i) \propto  \sum_{\sigma_{\partial i}} \frac{\exp \left[ -\beta \, \sigma_i \left( \sum_{k \in \partial i} J_{ki} \sigma_k + h_i \right) \right]}{ \cosh \left[ \beta \left( \sum_{k \in \partial i} J_{ki} \sigma_k + h_i \right) \right]} \, \prod_{k\in\partial i } p(\sigma_k|-\s_i)
 \label{eq:update_ME_numeric}
\end{eqnarray}

It is possible then to design an algorithm, analogous to BP, that finds the fixed-point values of conditional cavity probabilities through the updates equations (\ref{eq:update_CME_numeric}). Then equations (\ref{eq:update_ME_numeric}) can be used to obtain all the $P(\sigma_i)$, and therefore all local magnetizations. We will call this method CME-BP in what follows.

\begin{figure}[htb]
\centering
\includegraphics[keepaspectratio=true,width=0.55\textwidth]{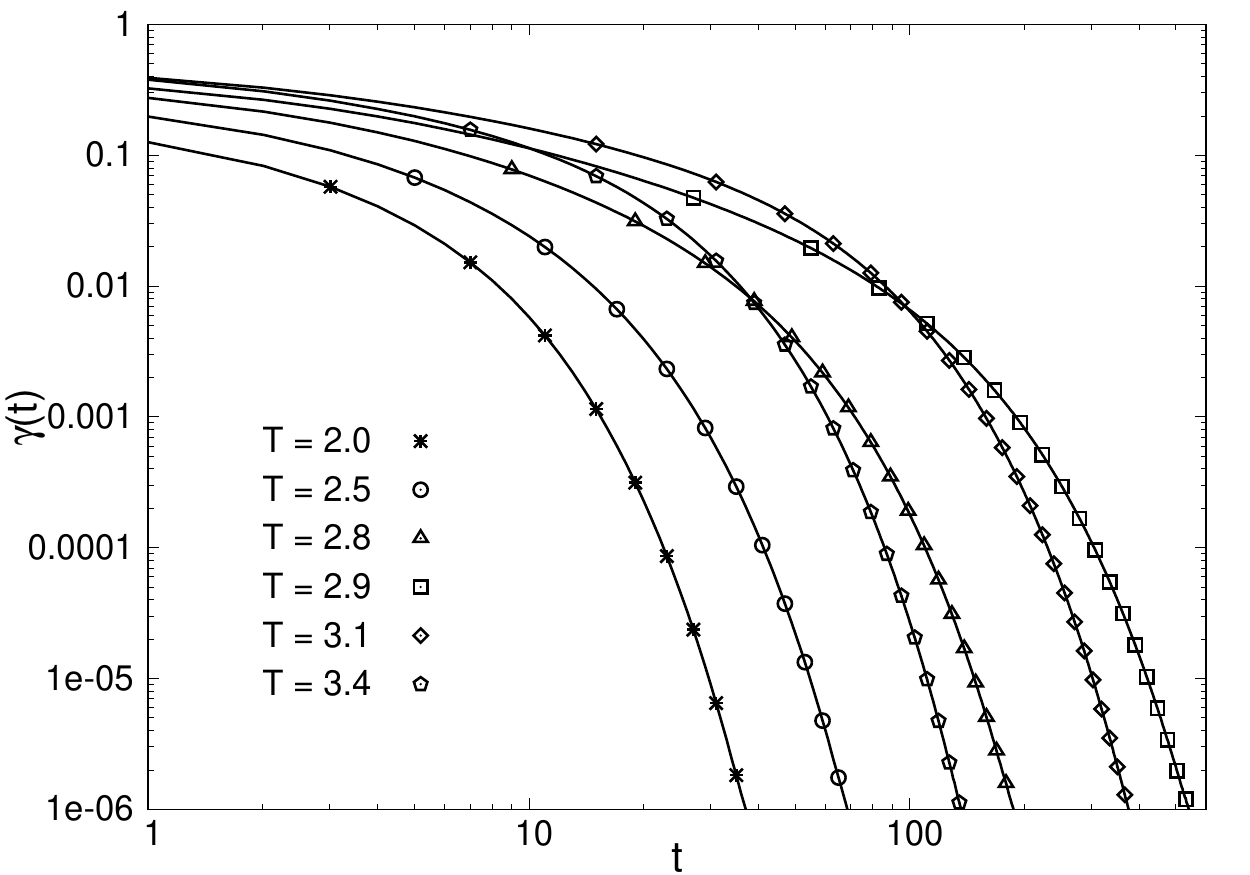}
\caption{Comparison between CME and CME-BP in a single Erdos-Renyi graph with mean connectivity $\langle c \rangle = 3$ and size $N=1004$. Figure shows the time dependence of the local error defined in (\ref{eq:local_error_CMEBP}) for several temperatures. All CME's calculations began with a fully magnetized system in contact with a heath bath. All CME-BP's runs started with all conditional cavity probabilities $p(\sigma_i = 1 \mid \sigma_j)$ set to one, and all $p(\sigma_i = -1 \mid \sigma_j)$ set to zero. Convergence parameter of the algorithm (see equation (\ref{eq:convergence_parameter})) was chosen as $\epsilon_{s} = 10^{-11}$.}
\label{fig:CME_vs_CMEBP_single_RGER}
\end{figure}

Figure (\ref{fig:CME_vs_CMEBP_single_RGER}) compares the results of CME-BP with the original dynamics obtained from numerical integration of original CME, $i.e$, of equations (\ref{eq:localMEfact}) and (\ref{eq:CME}), in a single Erdos-Renyi graph with mean connectivity $\langle c \rangle = 3$. We can define a local error like the one in (\ref{eq:local_error}) in order to quantitatively measure the agreement between CME-BP and original CME:

\begin{equation}
\gamma(t) = \sqrt{\frac{1}{N} \sum_{i} \left( m_{i}^{CME}(t) - m_{i}^{CME-BP} \right)^{2}}
 \label{eq:local_error_CMEBP}
\end{equation}

 Is important to notice that the error $\gamma(t)$ depends on time only through original CME's local magnetization: $m_{i}^{CME}(t)$, and that it goes to zero for all temperatures, which means that the solution of (\ref{eq:update_CME_numeric}) effectively corresponds to steady-state of (\ref{eq:CME}).
 
Of course, a fixed-point algorithm that uses (\ref{eq:update_CME_numeric}) to obtain stationary conditional cavity probabilities needs some convergence condition. Similarly as what is commonly done with standard BP, we defined the parameter:

\begin{equation}
\epsilon = \max_{\lbrace i, j, \sigma_i, \sigma_j \rbrace} \left[ \frac{\Delta p(\sigma_i \mid \sigma_j)}{p(\sigma_i \mid \sigma_j)} \right]
 \label{eq:convergence_parameter}
\end{equation}
where $\Delta p(\sigma_i \mid \sigma_j)$ is the change of $p(\sigma_i \mid \sigma_j)$ in one iteration of the algorithm. We stopped iterating when $\epsilon < \epsilon_s$, with $\epsilon_s$ some small positive real number.

\begin{figure}[htb]
\centering
\includegraphics[keepaspectratio=true,width=0.45\textwidth]{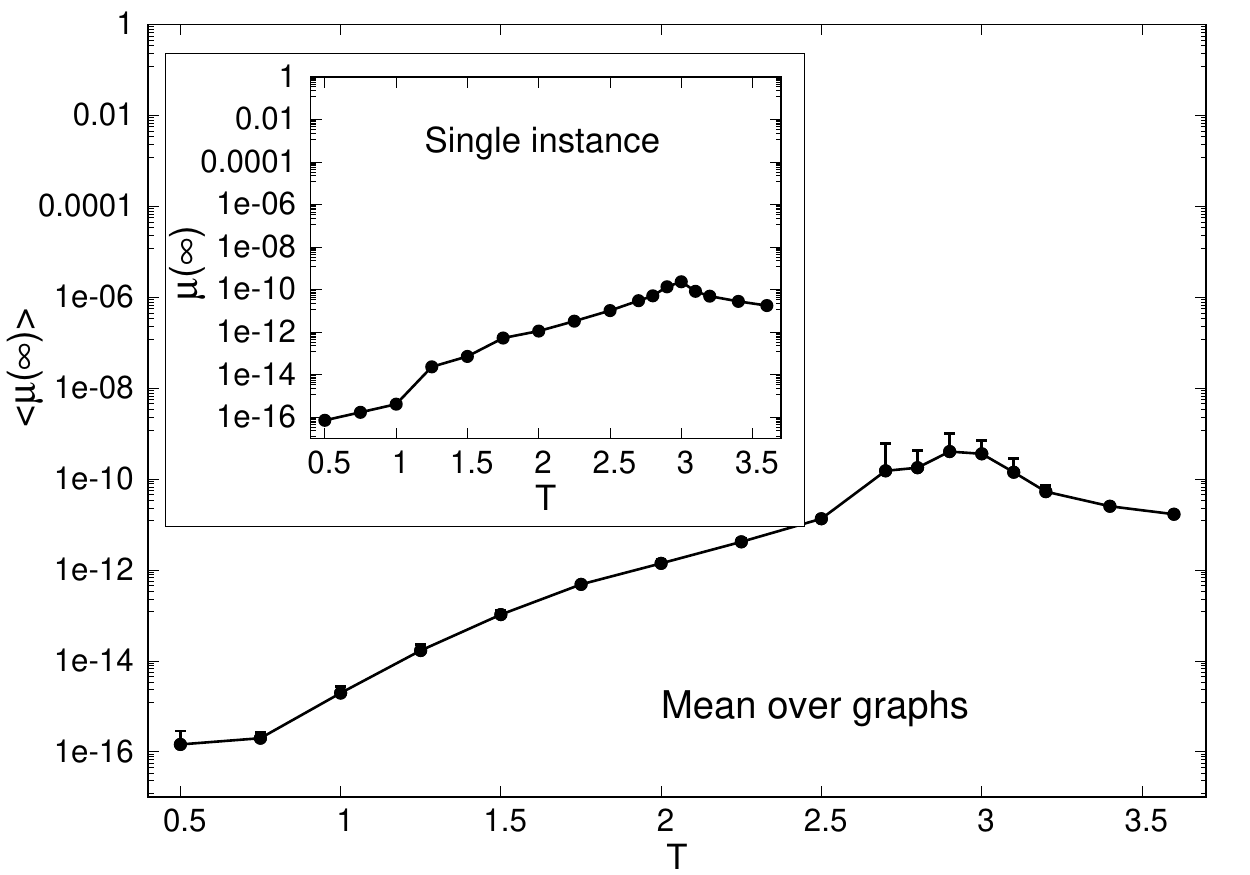}
\includegraphics[keepaspectratio=true,width=0.45\textwidth]{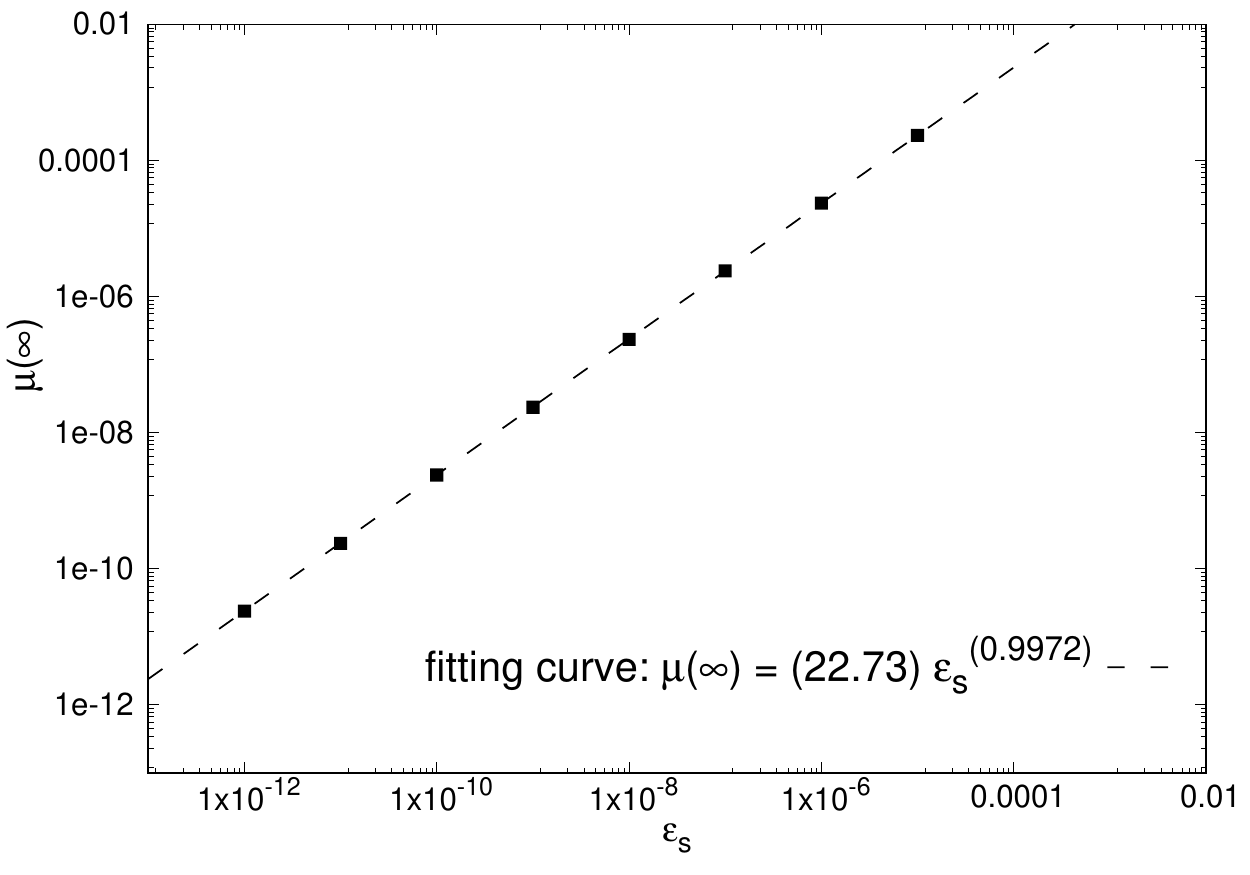}
\caption{Comparison between CME-BP and BP in Erdos-Renyi graphs with mean connectivity $\langle c \rangle=3$ and size $N=1004$. All CME-BP's runs started with all conditional cavity probabilities $p(\sigma_i = 1 \mid \sigma_j)$ set to one, and all $p(\sigma_i = 0 \mid \sigma_j)$ set to zero. Similarly, all BP's runs started with all messages $m_{i \rightarrow j}(\sigma_i = 1)$ set to one, and all $m_{i \rightarrow j}(\sigma_i = -1)$ set to zero. (left) Local error between local magnetizations corresponding to algorithms fixed-points. The inset shows the error in (\ref{eq:local_error_CMEBP_BP}) for several temperatures, for a single Erdos-Renyi graph, and main panel shows an average over different graphs of the same error. Error bars are drawn upwards from the point representing the average. The number of graphs was between $s = 20$ and $s = 100$. Convergence parameter of both algorithms (see equation (\ref{eq:convergence_parameter})) was chosen as $\epsilon_{s} = 10^{-11}$ (right) Dependence on convergence parameter $\epsilon_s$ of the error near $T_c^{ER} \approx 2.89$. Each point was computed using the same convergence parameter for both algorithms. All calculations were performed with $T=3.0$. Data was fitted to the curve $\mu(\infty) = a \, \epsilon_s^{b}$, with $a = 22.73 \pm 0.02$ and $b = 0.9972 \pm 0.0001$}
\label{fig:CMEBP_vs_BP_RGER}
\end{figure}

We are now ready to compare CME-BP with standard BP. The inset of figure (\ref{fig:CMEBP_vs_BP_RGER} (left)) compares the local magnetizations corresponding to fixed-points of BP and CME-BP. It shows, for a single Erdos-Renyi graph, the temperature dependence of the error:

\begin{equation}
\mu(\infty) = \sqrt{\frac{1}{N} \sum_{i} \left( m_{i}^{CME-BP} - m_{i}^{BP} \right)^{2}}
 \label{eq:local_error_CMEBP_BP}
\end{equation}

Main panel of the same figure displays the average  of that error over several Erdos-Renyi graphs with the same mean connectivity. In both graphics, error is small for all temperatures, and has a maximum near $T_{c}^{ER} \approx 2.89$. Figure (\ref{fig:CMEBP_vs_BP_RGER} right) shows that even under the influence of criticality, $\mu(\infty)$ goes to zero when the accuracy of the computations related to BP and CME-BP is improved by decreasing the convergence parameter $\epsilon_s$.

Summarizing, figure (\ref{fig:CMEBP_vs_BP_RGER}) shows a extremely good agreement between the equilibrium magnetizations predicted by BP and the stationary magnetizations of CME. Although, as we already discussed, there are significant theoretical differences, numerical similarities are a motivation for finding stronger analytical connections between both approaches.
In the next paragraph we establish this connection for all dynamics leading to thermal equilibrium.

\subsection{The CME steady state and the equilibrium Cavity Method.}

There is a close relation between the cavity distributions $\mu$ and the exact equilibrium conditional probabilities. For a diluted network the pair {\it equilibrium distribution} can be
written as:
\begin{equation}
 P(\s_i,\s_j) = \frac{1}{\partition_{ij}} \exp[\beta J_{ij}\s_i\s_j] \mu_{i\rightarrow j}(\s_i)
 \mu_{j\rightarrow i}(\s_j)
\end{equation}
from here it immediate to obtain the exact conditional distribution
\begin{equation}
 P(\s_i|\s_j) = \dfrac{1}{m_{i\rightarrow j}(J_{ij},\s_j)}\exp[\beta J_{ij}\s_i\s_j] \mu_{i\rightarrow j}(\s_i)
 \label{eqn:conditional_and_cavity_exact}
\end{equation}
This is rigourously true only for a tree-like network. In order to prove the equivalence of equilibrium CME with the exact result it is convenient to parametrize the CME distribution as \footnote{This is a general expression if $U_{ij}$ and $\mu'_{i\rightarrow j}(\s_i)$ are free parameters.}:

\begin{equation}
 p(\s_i|\s_j) = \dfrac{1}{m'_{i\rightarrow j}(U_{ij},\s_j)}\exp[\beta U_{ij}\s_i\s_j] \mu'_{i\rightarrow j}(\s_i)
 \label{eqn:conditional_parametrization}
\end{equation}
and show that the stationarity condition of the CME implies $U_{ij} = J_{ij}$ and also that $\mu'_{i\rightarrow j}(\s_i)$ satisfies \eqref{eqn:exact_cavity}. This will prove at once that 
the equilibrium CME solution is exactly \eqref{eqn:conditional_and_cavity_exact} and that the stationarity condition implies the equilibrium cavity equations.

Begining with the steady state condition $\frac{dp(\s_i|\s_j)}{dt}=0$ for the CME we get that the solutions satisfying detailed balance must obey:
\begin{equation}
\dfrac{p(\sigma_i | \sigma_j)}{p(-\sigma_i | \sigma_j)} =  \sum_{\sigma_{\partial i\setminus j}} \frac{r_{i}(-\sigma_i,\sigma_{\partial i})}{r_{i}(\sigma_i,\sigma_{\partial i})} \, \prod_{k\in\partial i \setminus j } p(\sigma_k|-\s_i)
 \label{eqn:conditional_prob_cavity}
\end{equation}
It is known that many different dynamics lead to the same Boltzmann equilibrium distribution. Therefore, to proceed with the derivation without lost of generality, we will only assume the transition rates satisfy 
\begin{equation}
 \frac{r_{i}(-\sigma_i,\sigma_{\partial i})}{r_{i}(\sigma_i,\sigma_{\partial i})}= \frac{P_{eq}(\s_i,\s_{\partial i})}{P_{eq}(-\s_i,\s_{\partial i})}= \exp[2\beta \s_i (\sum_{k\in \partial i}J_{ik}\s_k+ h_i)]
 \label{eqn:detailed_balance_rates}
\end{equation}
which is a consequence of imposing detailed balance in the exact master equation and having as a target for convergence the Boltzmann distribution. The next step is to insert \eqref{eqn:detailed_balance_rates} and the parametrization \eqref{eqn:conditional_parametrization} for all the conditional distributions in \eqref{eqn:conditional_prob_cavity}:
\begin{equation}
\exp[2\beta U_{ij}\s_i\s_j] \dfrac{\mu'_{i\rightarrow j}(\s_i)}{\mu'_{i\rightarrow j}(-\s_i)} =  \exp[2\beta J_{ij}\s_i\s_j] 
\dfrac
{\exp[ \beta h_i \s_i]\prod_{k\in\partial i \setminus j } m'_{k\rightarrow i}(2J_{ik} - U_{ik},\s_i)}
{\exp[-\beta h_i \s_i]\prod_{k\in\partial i \setminus j } m'_{k\rightarrow i}(U_{ik},-\s_i)}
 \label{eqn:intermediate_step_1}
\end{equation}
The first thing to notice in \eqref{eqn:intermediate_step_1} is that evaluating for $\sigma_j=\pm 1$
we get immediatly $U_{ij}=J_{ij}$. Naturally, we will use this result also for the rest of the $U_{ik}$ and write:
\begin{equation}
\dfrac{\mu'_{i\rightarrow j}(\s_i)}{\mu'_{i\rightarrow j}(-\s_i)} =
\dfrac
{\exp[ \beta h_i \s_i]\prod_{k\in\partial i \setminus j } m'_{k\rightarrow i}(J_{ik},\s_i)}
{\exp[-\beta h_i \s_i]\prod_{k\in\partial i \setminus j } m'_{k\rightarrow i}(J_{ik},-\s_i)}
 \label{eqn:intermediate_step_2}
\end{equation}

Equation \eqref{eqn:intermediate_step_2} implies directly that the trial distributions $\mu'$ introduced in the parametrization \eqref{eqn:conditional_parametrization} satisfy the equilibrium
cavity equations \eqref{eqn:exact_cavity}. This completes the proof that the stationary distributions
for the CME in a tree-like network are the exact conditional distributions and that the stationarity
condition for the CME is equivalent to the equilibrium cavity equations.

\section{Conclusions}
In this work we have revisited the Cavity Master Equations (CME). From the set of local equations that defines the dynamics of the system through CME we first derive a single equation for the dynamics of macroscopic variables in continuos time. We show how this equation, indeed correctly describes the behaviour of MC simulations after a proper average over different graphs of the ensemble. This opns a new door to efficiently study the dynamics of macroscopic quantities in systems with discrete variables and continuos dynamics. Finally we explore the connection of the CME with known techniques of equilibrium Statistical Physics. We proved that, the stationarity condition of the CME translates into a BP-like equation very similar to the one that can be derived by a proper minimization of the Bethe Free Energy. Furthermore, if in addition to stationarity one assumes a convergence to a Boltzmann distribution, CME is equivalent to the Cavity Equations derived in equilibrium. These results connects the dynamics of the CME with the equilibrium properties of models without frustration and may be a starting point to further explore the role of more sophsiticated phenomena like the Replica Symmetry Breaking of spin glasses from the dynamical point of view.

\section*{Acknowledgements}

We will like to thank Pr. E. Aurell for useful comments and suggestions on a previous version of this manuscript. This project has received funding from the European Union’s Horizon 2020 research and innovation programme MSCA-RISE-2016 under grant agreement No. 734439 INFERNET.

\bibliographystyle{unsrt}
\bibliography{ref_cvme}

\section*{Appendix A}
In this appendix, we would give some insight on how the approximations done when deriving the average case equations work. We can start by showing that the averaging process we followed conduces to correct numerical results. In order to do so, we performed a sampling process that we describe now for the Cavity Master Equation only:

\vspace{10pt}
\begin{algorithmic}[1] \label{Sampling_pseudo}
    \State{Choose two positive integers $M$ and $n$, and a real number $t_{max}$}
    \State{Initialize the vector $\lbrace p_{\gamma}(\sigma' \mid \sigma) \rbrace$, with $\gamma = 1, 2, ..., M$}
    \While{$t < t_{max}$}
    \For{$\gamma = (1, ..., M)$}
    \For{$i = (1, ..., n)$}
    \State{Choose a set $\lbrace \gamma_{k \in \lbrace 1, ..., \gamma \rbrace}^{i} \rbrace$} using $Q(\lbrace \gamma_{k}^{i} \rbrace)$
    \State{For these $\lbrace \gamma_{k}^{i} \rbrace$ construct the corresponding vector $\lbrace p_{\gamma_k}^{i}(\sigma' \mid \sigma) \rbrace$}
    \State{Compute $\dot{p}^{i}_{\gamma}(\sigma'| \sigma)$ from (\ref{eq:CME}) and using $\lbrace p_{\gamma_k}^{i}(\sigma' \mid \sigma) \rbrace$}
    \EndFor
    \State{Set $\dot{p}_{\gamma}(\sigma'| \sigma)$ as the average of all $\dot{p}^{i}_{\gamma}(\sigma'| \sigma)$}
    \EndFor
    \State{Update $\lbrace p_{\gamma}(\sigma' \mid \sigma) \rbrace$}
    \State{Compute the average $p_{\lambda}(\sigma' \mid \sigma)$ of all $\lbrace p_{\gamma}(\sigma' \mid \sigma) \rbrace$ using $Q(\gamma)$}
    \EndWhile 
\end{algorithmic}

\begin{figure}[htb]
\centering
\includegraphics[keepaspectratio=true,width=0.55\textwidth]{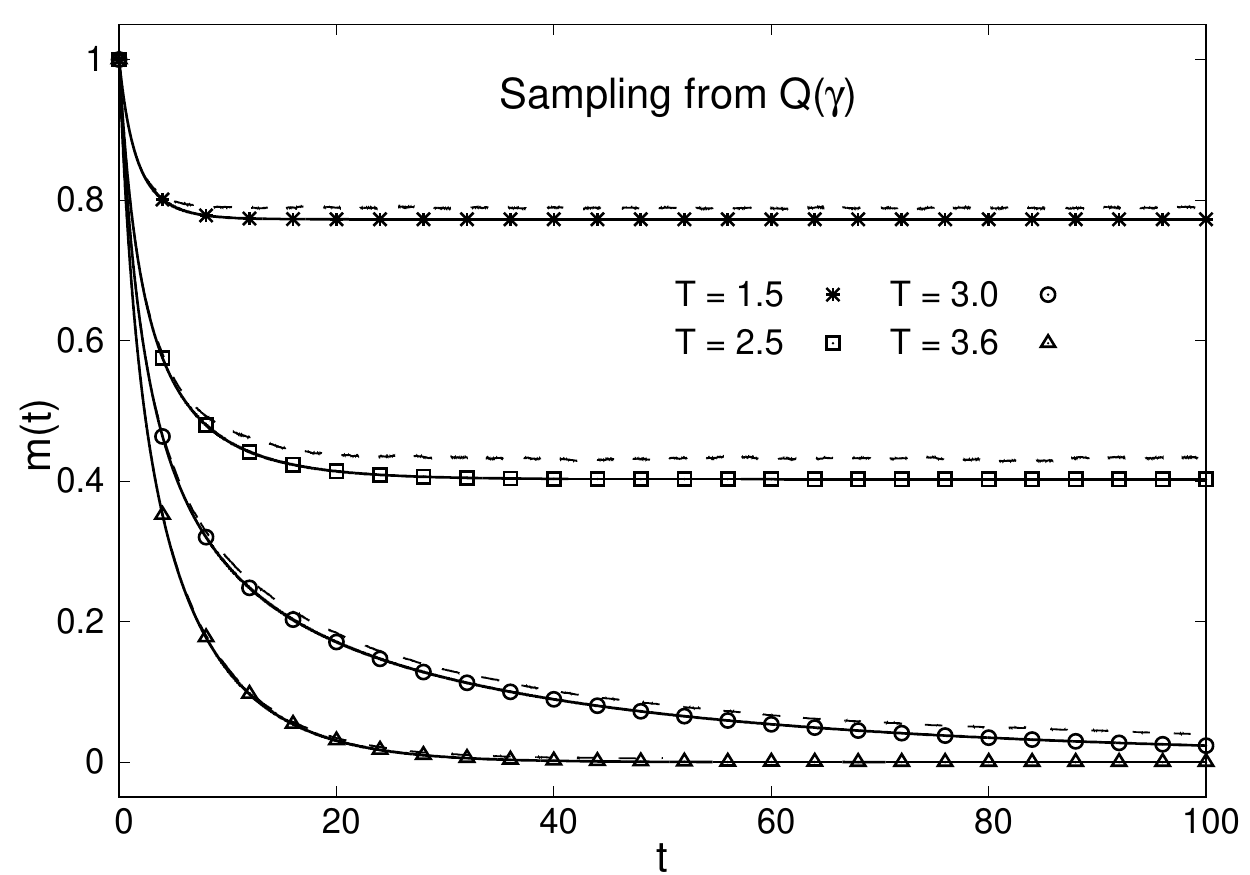}
\caption{Comparison between Sampled CME (dashed lines) and regular CME (lines a\
nd points), for calculations with initial magnetization $m_0=1.0$ in an Erdos-R\
enyi graph whose mean connectivity is $c=3$ ($T_c \approx 2.88$). We assumed $p\
_{\gamma > 50}(\sigma' \mid \sigma) = 1$. In regular CME, all calculations bega\
n with a fully magnetized in contact with a heath bath at a given temperature $\
T$. Curves are the average of calculations done for different graphs. System's \
size is in all cases $N=4000$. The number of graphs was $s=350$.}
\label{fig:RGER_average_sampling}
\end{figure}

Figure (\ref{fig:RGER_average_sampling}) compares the results of this scheme, that we called Sampled CME, with the ones of regular CME. In order to obtain observables as the magnetization, we need to follow an analogous procedure, but for the $P(\sigma_i)$ probabilities. Its description will be omitted here.

As can be seen, Sampled CME gives a good description of original equation. Note that when choosing the set $\lbrace \gamma_{k}^{i} \rbrace$ nothing stops us of selecting some very large $\gamma_k^{i}$. We need to make some approximation for large connectivity that allow us to implement our sampling procedure. In this case we assumed that $p_{\gamma}(\sigma' \mid \sigma) = 1$, for all $\gamma > \gamma_{max}$, with $\gamma_{max}$ being some positive integer.

At some point of the derivation, we have found differential equations for averaged probabilities $p_{\lambda}( \sigma_{i} \mid \sigma_{j})$ and $P_{\lambda}(\sigma_i)$. We could find a set of differential equation for all $p_{\gamma_i}( \sigma_{i} \mid \sigma_{j})$ and $P_{c_i}(\sigma_i)$ just by omitting the sums over $\gamma_i$ and $c_i$ in (\ref{eq:first_sum_CME_av_graph_2}) and (\ref{eq:first_sum_ME_av_graph_2}), respectively:

 \begin{eqnarray}
 \lefteqn{\dot{p}_{\gamma_j}(\sigma_{j}| \sigma_i)=- \dfrac{\alpha}{2} \lbrace p_{\gamma_j}(\sigma_{j} \mid \sigma_{i}) -  p_{\gamma_j}(-\sigma_{j} \mid \sigma_{i}) \rbrace + \dfrac{\alpha}{2} \sigma_j \sum_{\sigma_{\partial j \setminus i}} \tanh(\beta J \sum_{k \in \partial j} \sigma_k) \times } \nonumber  \\ 
 && \times \left[ \displaystyle \prod_{k \in \partial j \setminus i} p_{\lambda}( \sigma_{k} \mid \sigma_{j}) \; p_{\gamma_j}(\sigma_{j} \mid \sigma_{i}) + \displaystyle \prod_{k \in \partial j \setminus i} p_{\lambda}( \sigma_{k} \mid - \sigma_{j}) \; p_{\gamma_j}(-\sigma_{j} \mid \sigma_{i})  \right]
\label{eq:first_sum_CME_av_graph_1_for_p_gamma} \\
\nonumber
 \lefteqn{\dot{P}_{c_i}(\sigma_{i})=- \dfrac{\alpha}{2} \lbrace P_{c_i}(\sigma_{i}) -  P_{c_i}(-\sigma_{i}) \rbrace + \dfrac{\alpha}{2} \sigma_i \sum_{\sigma_{\partial i}} \tanh(\beta J \sum_{k \in \partial i} \sigma_k) \times }\\ 
 && \times \left[ \displaystyle \prod_{k \in \partial i} p_{\lambda}( \sigma_{k} \mid \sigma_{i}) \; P_{c_i}(\sigma_{i}) + \displaystyle \prod_{k \in \partial i} p_{\lambda}( \sigma_{k} \mid - \sigma_{i}) \; P_{c_i}(-\sigma_{i})  \right]
\label{eq:first_sum_ME_av_graph_1_for_c_i}
\end{eqnarray}

We can use equations (\ref{eq:first_sum_CME_av_graph_2}),  (\ref{eq:first_sum_CME_av_graph_1_for_p_gamma}), (\ref{eq:first_sum_ME_av_graph_2}) and (\ref{eq:first_sum_ME_av_graph_1_for_c_i}) to get some numeric results. In what follows, we describe the procedure we followed in the case of Cavity Master Equation. Master Equation can be treated, again, by analogy.

\vspace{10pt}
\begin{algorithmic}[1] \label{Summing_mean_pseudo}
    \State{Choose a positive integer $M$ and a real number $t_{max}$}
    \State{Initialize the vector $\lbrace p_{\gamma}(\sigma' \mid \sigma) \rbrace$, with $\gamma = 1, 2, ..., M$}
    \While{$t < t_{max}$}
    \For{$\gamma = (1, ..., M)$}
    \State{Compute $\dot{p}_{\gamma}(\sigma'| \sigma)$ from (\ref{eq:first_sum_CME_av_graph_1_for_p_gamma})}
    \EndFor
    \State{Update $\lbrace p_{\gamma}(\sigma' \mid \sigma) \rbrace$}
    \State{Compute the average $p_{\lambda}(\sigma' \mid \sigma)$ of all $\lbrace p_{\gamma}(\sigma' \mid \sigma) \rbrace$ using $Q(\gamma)$}
    \EndWhile 
\end{algorithmic}

\begin{figure}[htb]
\centering
\includegraphics[keepaspectratio=true,width=0.55\textwidth]{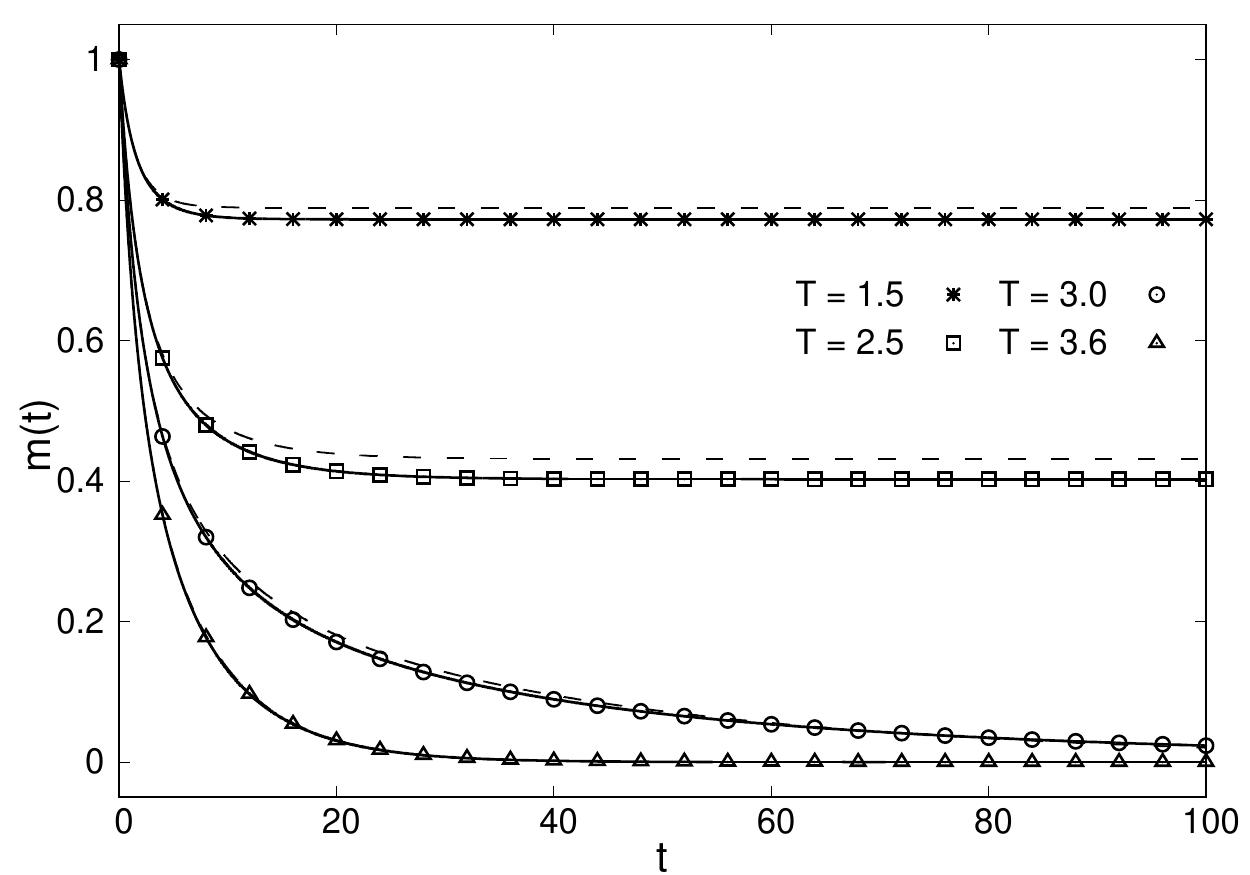}
\caption{Comparison between $1^{st}$ average CME (dashed lines) and regular CME (lines and points), for calculations with initial magnetization $m_0=1.0$ in an Erdos-Renyi graph whose mean connectivity is $c=3$ ($T_c \approx 2.88$). In both graphics the sums in (\ref{eq:first_sum_CME_av_graph_2}) and (\ref{eq:first_sum_ME_av_graph_2}) were approximately computed using the first $50$ terms. In regular CME, all calculations began with a fully magnetized system in contact with a heath bath at a given temperature $T$. Curves are the average of calculations done for different graphs. System's size is in all cases $N=4000$. The number of graphs was $s=350$.}
\label{fig:RGER_average_summing_mean}
\end{figure}

Figure (\ref{fig:RGER_average_summing_mean}) compares the results obtained this way, that we called $1^{st}$ Average CME, with the ones of regular CME. As can be seen, $1^{st}$ Average CME gives a good description of original equation.

\end{document}